\documentclass[showpacs,superscriptaddress,floatfix,pra,twocolumn,10pt]{revtex4}
\usepackage{graphicx}
\usepackage{amsmath,amssymb}
\usepackage{bm}

\input{tcilatex}

\begin{document}

\title{Effective spin-1 Heisenberg chain in coupled cavities}
\author{Ling Zhou, Wei-Bin Yan, Xin-Yu Zhao}

\begin{abstract}
A coupled array of $N$ identical cavities, each of which contains a
five-level atom is investigated. The results show that the atoms via the
exchange of virtual photons can be effectively equal to spin 1 Heisenberg
model under certain conditions. By tuning the laser fields, the parameters
of the effective Hamiltonian can be controlled individually.
\end{abstract}

\maketitle

\address{School of physics and optoelectronic technology, Dalian University
of Technology, Dalian 116024, P.R.China}

\section{Introduction}

Spin chain has been played an important role in quantum information field as
well as in condensed matter physics. The interaction between on site spin
can offer us entanglement in solid and realistic way [1-7]. It has been
found that spin chain with an open boundary condition can be applied in
quantum communication \cite{bose}, which can translate information with high
fidelity from one end to another. Perfect state transfer can be realized by
changing the interaction between the qubits in spin networks \cite{miche,
woj}.

However, because of the microscopic properties of solid-state materials, it
is very hard to address individual spins while it is the prerequisite for
quantum information processing. Single spin addressability can also be very
helpful to obtain deeper and more detail insight into condensed matter
physics. In order to do this, it has been show that the arrays of Josephson
junctions \cite{JUC}, quantum dots \cite{DL}, optical lattices \cite{lattice}%
, can provide effective spin-chain Hamiltonian where spin-coupling constants
can be controlled. Recently under active investigation is to use the array
of coupled cavities, which are ideally suited to addressing individual spins
[14-23]. Intense interest has arisen from the demonstration that a
polaritonic Mott transition and a Bose-Hubbard interaction can be generated
in these structures \cite{ank,hartmann,green}. Hartmann \cite{hartmann2}
have shown that single atoms in interacting cavities that are operated in a
strong coupling regime can form a Heisenberg spin $\frac{1}{2}$ Hamiltonian
in which all parameters of the effective Hamiltonian can be tuned
individually.

However, people are increasedly interested in the multilevel systems. A lots
of papers treat with multilevel systems in different areas of physics such
as in condensed matter physics\cite{cond1,cond2}, statistical mechanics \cite%
{yip, pire}, as well as in quantum information \cite{quant1,quant2,bruss}.
Now multilevel systems can be considered as a important field. In this
paper, we consider a coupled array of $N$ identical cavities, each of which
contain a five-level atom. We show that under large detuning case the atoms
via the exchange of virtual photons can be effectively equal to spin 1
model. We use atomic bare basis as spin-level. It should be easy to
manipulate the individual atom when we need to project measure etc. on the
individual qutrit.

\section{\protect\bigskip Model and the effective Hamiltonian}

\bigskip We consider an array of cavities which are coupled via exchange of
photons. Each of the cavities contains one five-level atom. The atomic
configuration is shown in Fig.1. The three long-lived levels $|a\rangle $, $%
|b\rangle $, $|c\rangle $ represent the three spin states. The cavity mode
couples to the transitions $|d\rangle \leftrightarrow |a\rangle $, $%
|d\rangle \leftrightarrow |b\rangle $, $|e\rangle \leftrightarrow |b\rangle $%
, and $|e\rangle \leftrightarrow |c\rangle $ while four lasers drive the
atom with Rabi frequencies $\Omega _{i}$ $(i=1,2,3,4)$, respectively. The
same atomic configurations has been used in \cite{cho3} for addressing
individual atoms in optical lattices with standing-wave driving fields. In
the interaction picture, the Hamiltonian reads\bigskip

\begin{eqnarray}
H &=&\sum_{j=1}^{N}[(g_{1}a_{j}|d_{j}\rangle \langle a_{j}|+\Omega
_{1}|d_{j}\rangle \langle b_{j}|)e^{i\Delta _{1}t} \\
&&+(g_{2}a_{j}|d_{j}\rangle \langle b_{j}|+\Omega _{2}|d_{j}\rangle \langle
a|)e^{i\Delta _{2}t}  \notag \\
&&+(g_{3}a_{j}|e_{j}\rangle \langle b_{j}|+\Omega _{3}|e_{j}\rangle \langle
c_{j}|)e^{i\Delta _{3}t}  \notag \\
&&+(g_{4}a_{j}|e_{j}\rangle \langle c_{j}|+\Omega _{4}|e_{j}\rangle \langle
b_{j}|)e^{i\Delta _{4}t}+h.c.]  \notag \\
&&+\sum_{j=1}^{N}J(a_{j}^{\dagger }a_{j+1}+a_{j}a_{j+1}^{\dagger }),  \notag
\end{eqnarray}%
where $g_{i}$ is the coupling strength between the cavity and the atom, and $%
\Delta _{i}$ express the detuning shown in Fig.1, and $a_{j}$ means the
annihilation operator of the cavity field. Here, for simplicity, we assume
that the coupling strength $g_{i}$ and laser Rabi frequency $\Omega _{i}$ do
not change with cavity sequence $j$, which means that the identical atoms
are in the identical cavities and driven by identical laser fields. The
first sum denote the interaction within the cavities, and the second sum is
the inter-cavity hopping between nearest neighbor with hopping rate $J$. Now
we consider the large detuning case with $|\Delta _{i}|$ $\gg |g_{i}|$,$%
|\Omega _{i}|$. In addition, in order to avoid undesired atomic transitions,
we need the following inequalities $\Delta _{2}-\Delta _{1}\gg \{\frac{%
\Omega _{1}\Omega _{2}}{\Delta _{1}},\frac{\Omega _{1}g_{2}}{\Delta _{1}},%
\frac{g_{1}\Omega _{2}}{\Delta _{1}},\frac{g_{1}g_{2}}{\Delta _{1}}\}$, $%
\Delta _{4}-\Delta _{3}\gg \{\frac{\Omega _{3}\Omega _{4}}{\Delta _{3}},%
\frac{\Omega _{3}g_{4}}{\Delta _{3}}\frac{g_{3}\Omega _{4}}{\Delta _{3}}%
\frac{g_{3}g_{4}}{\Delta _{3}}\}$. We adiabatic eliminate atomic excited
state $|d\rangle $ and $|e\rangle $ and have 
\begin{figure}[tbp]
\includegraphics[width=8cm, height=5cm]{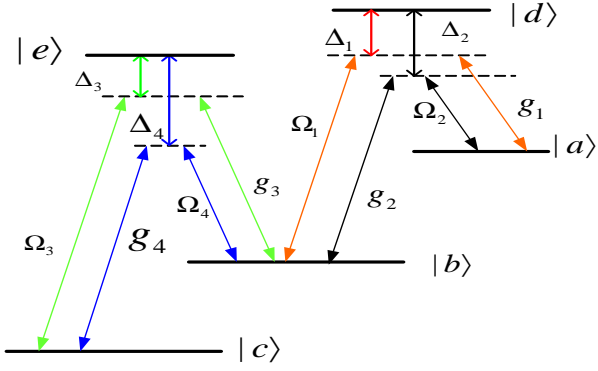}
\caption{(Color online) The configuration of the five-level atom. The three
long-lived levels $|a\rangle $, $|b\rangle $, $|c\rangle $ represent the
three spin states. The cavity mode couples to the transtions $|d\rangle
\leftrightarrow |a\rangle $, $|d\rangle \leftrightarrow |b\rangle $, $%
|e\rangle \leftrightarrow |b\rangle $, and $|e\rangle \leftrightarrow
|c\rangle $ while four lasers drive the atom with Rabi frequencies $\Omega
_{i}$ $(i=1,2,3,4)$, respectively. }
\end{figure}
\ \ \ 
\begin{eqnarray}
H &=&-\sum_{j=1}^{N}[\frac{g_{1}^{2}}{\Delta _{1}}|a_{j}\rangle \langle
a_{j}|+(\frac{g_{2}^{2}}{\Delta _{2}}+\frac{g_{3}^{2}}{\Delta _{3}}%
)|b_{j}\rangle \langle b_{j}|+\frac{g_{4}^{2}}{\Delta _{4}}|c_{j}\rangle
\langle c_{j}|]a_{j}^{\dagger }a_{j}  \notag \\
&&-\sum_{j=1}^{N}[\frac{\Omega _{2}^{2}}{\Delta _{2}}|a_{j}\rangle \langle
a_{j}|+(\frac{\Omega _{1}^{2}}{\Delta _{1}}+\frac{\Omega _{4}^{2}}{\Delta
_{4}})|b_{j}\rangle \langle b_{j}|+\frac{\Omega _{3}^{2}}{\Delta _{3}}%
|c_{j}\rangle \langle c_{j}|]  \notag \\
&&-\sum_{j=1}^{N}[\frac{\Omega _{1}g_{1}}{\Delta _{1}}a_{j}|b_{j}\rangle
\langle a_{j}|+\frac{\Omega _{2}g_{2}}{\Delta _{2}}a_{j}|a_{j}\rangle
\langle b_{j}|  \notag \\
&&+\frac{\Omega _{3}g_{3}}{\Delta _{3}}a_{j}|c_{j}\rangle \langle b_{j}|+%
\frac{\Omega _{4}g_{4}}{\Delta _{4}}a_{j}|b_{j}\rangle \langle c_{j}|+h.c.] 
\notag \\
&&+\sum_{j=1}^{N}J(a_{j}^{\dagger }a_{j+1}+a_{j}a_{j+1}^{\dagger }).
\end{eqnarray}%
We assume 
\begin{equation}
\frac{g_{1}^{2}}{\Delta _{1}}=\frac{g_{2}^{2}}{\Delta _{2}}+\frac{g_{3}^{2}}{%
\Delta _{3}}=\frac{g_{4}^{2}}{\Delta _{4}}.
\end{equation}%
The effective Hamiltonian can be classified as 
\begin{equation}
H=H_{0}+H_{1}
\end{equation}%
with%
\begin{eqnarray*}
H_{0} &=&-\sum_{j=1}^{N}[\frac{g_{1}^{2}}{\Delta _{1}}a_{j}^{\dagger
}a_{j}-J(a_{j}^{\dagger }a_{j+1}+a_{j}a_{j+1}^{\dagger })] \\
&&-\sum_{j=1}^{N}[\frac{\Omega _{2}^{2}}{\Delta _{2}}|a_{j}\rangle \langle
a_{j}|+(\frac{\Omega _{1}^{2}}{\Delta _{1}}+\frac{\Omega _{4}^{2}}{\Delta
_{4}})|b_{j}\rangle \langle b_{j}|+\frac{\Omega _{3}^{2}}{\Delta _{3}}%
|c_{j}\rangle \langle c_{j}|],
\end{eqnarray*}

\begin{eqnarray}
H_{1} &=&-\sum_{j=1}^{N}[\frac{\Omega _{1}g_{1}}{\Delta _{1}}%
a_{j}|b_{j}\rangle \langle a_{j}|+\frac{\Omega _{2}g_{2}}{\Delta _{2}}%
a_{j}|a_{j}\rangle \langle b_{j}|  \notag \\
&&+\frac{\Omega _{3}g_{3}}{\Delta _{3}}a_{j}|c_{j}\rangle \langle b_{j}|+%
\frac{\Omega _{4}g_{4}}{\Delta _{4}}a_{j}|b_{j}\rangle \langle c_{j}|+h.c.].
\end{eqnarray}%
Noting that the every term in $H_{0}$ commutes each other, when we perform
unitary transformation we can do it separately. Diagonalize the
cavity-hopping terms by employing a Fourier transformed basis as $a_{j}=%
\frac{1}{\sqrt{N}}\sum_{k=1}^{N}F_{jk}b_{k}$, where $F_{jk}$ $=exp(-i\frac{%
2\pi }{N}jk)$ and $\sum_{j=1}^{N}F_{jk}F_{jl}^{\ast }=N\delta _{kl\text{.}}$
The diagonalized form reads $J\sum_{j}(a_{j}^{\dag
}a_{j+1}+a_{j}a_{j+1}^{\dag })=\sum_{j}$ $\nu _{j}b_{j}^{\dag }b_{j}$ with $%
\nu _{j}=2J\cos (\frac{2\pi }{N}j)$. Now, we goes into the new frame
rotating with $H_{0}$, under the conditions%
\begin{eqnarray}
\frac{\Omega _{1}^{2}}{\Delta _{1}}+\frac{\Omega _{4}^{2}}{\Delta _{4}} &=&%
\frac{1}{2}(\frac{\Omega _{2}^{2}}{\Delta _{2}}+\frac{\Omega _{3}^{2}}{%
\Delta _{3}}), \\
\frac{\Omega _{1}g_{1}}{\Delta _{1}} &=&\frac{\Omega _{3}g_{3}}{\Delta _{3}},%
\frac{\Omega _{2}g_{2}}{\Delta _{2}}=\frac{\Omega _{4}g_{4}}{\Delta _{4}}, 
\notag
\end{eqnarray}%
we have the Hamiltonian 
\begin{eqnarray}
H &=&-\sum_{j=1}^{N}\{[\frac{\Omega _{1}g_{1}}{\Delta _{1}}(|b_{j}\rangle
\langle a_{j}|+|c_{j}\rangle \langle b_{j}|)e^{i\mu _{+}t} \\
&&+\frac{\Omega _{2}g_{2}}{\Delta _{2}}(|a_{j}\rangle \langle
b_{j}|+|b_{j}\rangle \langle c_{j}|)e^{i\mu _{\_}t}]  \notag \\
&&e^{-i\nu _{j}t}F_{jk}b_{k}+h.c.\}  \notag
\end{eqnarray}%
with $\mu _{\pm }=\frac{g_{1}^{2}}{\Delta _{1}}\pm \frac{1}{2}(\frac{\Omega
_{2}^{2}}{\Delta _{2}}-\frac{\Omega _{3}^{2}}{\Delta _{3}})$. We define that
the eigenstates of $S_{z}$ for $S=1$ are atomic bare level $|a\rangle $, $%
|b\rangle $, $|c\rangle $; therefore $S_{i+}=S_{ix}+iS_{iy}=\sqrt{2}%
(|a\rangle \langle b|+|b\rangle \langle c|)$, $S_{i-}=S_{ix}-iS_{iy}=\sqrt{2}%
(|b\rangle \langle a|+|c\rangle \langle b|)$. Using the spin operator, we
rewrite the Hamiltonian Eq.(7) as 
\begin{eqnarray}
H &=&-\sum_{j,k=1}^{N}[(\frac{\Omega _{1}g_{1}}{\sqrt{2}\Delta _{1}}%
S_{j}^{-}e^{i\mu _{+}t}  \notag \\
&&+\frac{\Omega _{2}g_{2}}{\sqrt{2}\Delta _{2}}S_{j}^{+}e^{i\mu
_{-}t})e^{-i\nu _{k}t}F_{jk}b_{k}+h.c.].
\end{eqnarray}%
If the Rabi frequency \{$\frac{\Omega _{1}g_{1}}{\sqrt{2}\Delta _{1}}$,$%
\frac{\Omega _{2}g_{2}}{\sqrt{2}\Delta _{2}}$\}$\ll \{|\mu _{+}-\nu
_{k}|,|\mu _{-}-\nu _{k}|\}$, this allows us to make use of the adiabatic
elimination once more. Considering the subspace without real photons, we
deduce the effective Hamiltonian as%
\begin{equation}
H=\sum_{j=1}^{N}[\frac{\Omega _{1}^{2}g_{1}^{2}}{2\Delta _{1}^{2}}\frac{1}{%
\mu _{+}-\nu _{k}}S_{j}^{-}S_{l}^{+}+\frac{\Omega _{2}^{2}g_{2}^{2}}{2\Delta
_{2}^{2}}\frac{1}{\mu _{-}-\nu _{k}}S_{j}^{+}S_{l}^{-}]F_{jk}F_{lk}^{\ast }.
\end{equation}%
In view of $\nu _{k}$ $=J(F_{1k}+F_{1k}^{\ast })$, we expand $\frac{%
F_{jk}F_{lk}^{\ast }}{\mu _{\pm }-\nu _{k}}$ as%
\begin{equation}
\frac{F_{jk}F_{lk}^{\ast }}{\mu _{\pm }-\nu _{k}}=\frac{1}{\mu _{\pm }}%
(F_{jk}F_{lk}^{\ast }+\frac{J}{\mu _{\pm }}F_{j+1k}F_{lk}^{\ast }+\frac{J}{%
\mu _{\pm }}F_{jk}F_{l+1k}^{\ast }).
\end{equation}%
This expand demand the condition $J\prec \mu _{\pm }$. Substitute the
relation Eq.(10) into Eq.(9), we finally obtain the effective Hamiltonian 
\begin{figure}[tp]
\includegraphics[width=7cm, height=5cm]{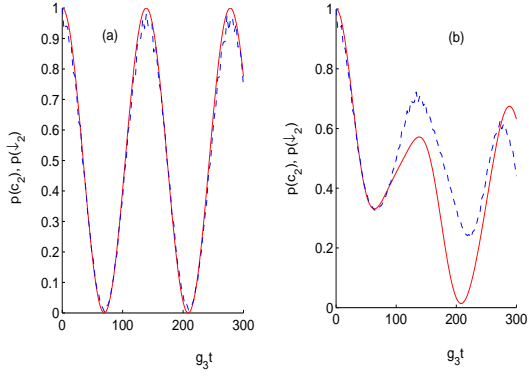}
\caption{(Color online) The comparison of the probility $p(c_{2})$ [ $%
p(\downarrow _{2})]$ between the Hamiltonian Eq.(2)(blue-dashed line) with
the effective spin chain Eq.(11)(red-solid line). (a): the initial atomic
state is $|b_{1},c_{2}\rangle $ . (b): the initial atomic state is $\frac{1}{%
2}(|a_{1}\rangle \langle a_{1}|+|b_{1}\rangle \langle b_{1}|)\otimes
|c_{2}\rangle \langle c_{2}|$ ($\frac{1}{2}(|\uparrow _{1}\rangle \langle
\uparrow _{1}|+|\rightarrow _{1}\rangle \langle \rightarrow _{1}|)\otimes
|\downarrow _{2}\rangle \langle \downarrow _{2}|$ for spin chain Eq.(11).
For all plots, the parameters are $g_{1}=g_{2}=g_{4}=1$, $g_{3}=1/2$, $%
\Delta _{1}=\Delta _{4}=40$,$\Delta _{3}=20$, $\Delta _{2}=80$, $\Omega
_{1}=\Omega _{2}=\Omega _{3}=10$, $\Omega _{4}=5$, $J=0.5$.}
\end{figure}
\begin{equation}
H_{xy}=%
\sum_{j=1}^{N}A[S_{xj}^{2}+S_{yj}^{2}]+BS_{jz}+C(S_{xj}S_{xj+1}+S_{yj}S_{yj+1})
\end{equation}%
with 
\begin{eqnarray}
A &=&\frac{\Omega _{1}^{2}g_{1}^{2}}{2\Delta _{1}^{2}\mu _{+}}+\frac{\Omega
_{2}^{2}g_{2}^{2}}{2\Delta _{2}^{2}\mu _{-}},B=\frac{\Omega _{2}^{2}g_{2}^{2}%
}{2\Delta _{2}^{2}\mu _{-}}-\frac{\Omega _{1}^{2}g_{1}^{2}}{2\Delta
_{1}^{2}\mu _{+}},  \notag \\
C &=&\frac{\Omega _{1}^{2}g_{1}^{2}J}{\Delta _{1}^{2}\mu _{+}^{2}}+\frac{%
\Omega _{2}^{2}g_{2}^{2}J}{\Delta _{2}^{2}\mu _{-}^{2}}.
\end{eqnarray}%
We see clearly that it is spin 1 XY antiferromagnetic Hamiltonian ($C>0$).
Because it is spin 1, the term $S_{xj}^{2}+S_{yj}^{2}$ can not be omitted,
which is of essential importance in high-spin cases. Although individual
control of the coefficients $A$, $B$, $C$ is limited owing to their mutual
dependency, we still can change them; because $\{g_{i},\Delta _{i},\Omega
_{i}\}$ ($i=1...4$) meeting with the condition Eqs.(3),(6), we still have
seven variable so as to adjust the coefficients $A$, $B$, $C.$

To confirm the validity of our approximations, we numerically simulate the
dynamics generated by Hamiltonian Eq.(2) and compare it to the dynamics
generated by effective spin model Eq.(11).\ In Fig. 2a, we consider two
atoms in two cavities initially in the state $|b_{1},c_{2}\rangle $ (atom
one in $|b\rangle $ and atom two in $|c\rangle $ ) and calculate the
occupation probability $p(c_{2})$ (atom 2 in state $|c\rangle )$ which
corresponding to the probability of spin 2 to point down. Fig.2a shows the
time evolution of $p(c_{2})$ for Hamiltonian Eq.(2) and $p(\downarrow _{2})$
for effective spin mode Eq.(11). For our choice of the parameters, one can
easy check that they satisfy all kinds of approximation condition such as
Eqs.(3), (6) and the two times adiabatic elimination condition $|\Delta
_{i}|\gg $\{$|g_{i}|$,$|\Omega _{i}|$\},$\Delta _{2}-\Delta _{1}\gg \{\frac{%
\Omega _{1}\Omega _{2}}{\Delta _{1}},\frac{\Omega _{1}g_{2}}{\Delta _{1}},%
\frac{g_{1}\Omega _{2}}{\Delta _{1}},\frac{g_{1}g_{2}}{\Delta _{1}}\}$, $%
\Delta _{4}-\Delta _{3}\gg \{\frac{\Omega _{3}\Omega _{4}}{\Delta _{3}},%
\frac{\Omega _{3}g_{4}}{\Delta _{3}}\frac{g_{3}\Omega _{4}}{\Delta _{3}}%
\frac{g_{3}g_{4}}{\Delta _{3}}\}$, and $\{\frac{\Omega _{1}g_{1}}{\sqrt{2}%
\Delta _{1}}$,$\frac{\Omega _{2}g_{2}}{\sqrt{2}\Delta _{2}}\}\ll $\{$|\mu
_{+}-\nu _{k}|,|\mu _{-}-\nu _{k}|$\}. From the Fig. 2a, we know that the
interaction can be effectively equal to spin 1 Heisenberg chain because the
two curves almost merge into one. For the initial atomic state $%
|b_{1},c_{2}\rangle $, the population of two atoms just oscillate between
the two level, and level $|a\rangle $ do not participate the interaction so
that we have the reduced two-level system (level $|a\rangle $ has no
population).

\ Fig.2b plots the population $p(c_{2})$ for initial state $\frac{1}{2}%
(|a_{1}\rangle \langle a_{1}|+|b_{1}\rangle \langle b_{1}|)\otimes
|c_{2}\rangle $. Under this case, all the three levels participate the
interaction so that we can see the two steps oscillation (see red line). But
now, we have much discrepancy between Hamiltonian Eq. (2) and effective
Hamiltonian Eq. (11). The discrepancies come out of the higher order term in
the adiabatic elimination. Due to the two participators the excited level
and the middle level, the discrepancy between Hamiltonian Eq. (2) and
effective Hamiltonian Eq. (11) become larger.

Now, we will obtain effective $S_{Z}S_{Z}$ interaction. We still employ the
atomic level configuration but now we only need the two laser beams working
between $|e\rangle \leftrightarrow |c\rangle $ and $|d\rangle
\leftrightarrow |a\rangle $ shown in Fig.3. The Hamiltonian is as 
\begin{figure}[tp]
\includegraphics[width=7cm, height=5cm]{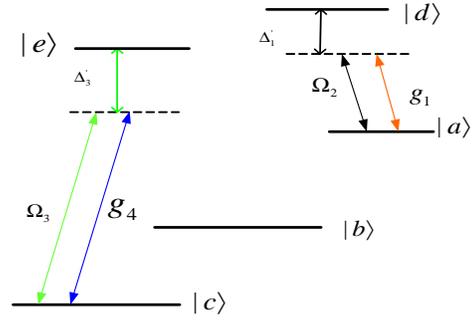}
\caption{(Color online) The illustration to get $S_{Z}S_{Z}$ interaction.
Two classical fields are deleted}
\end{figure}
\begin{eqnarray}
H &=&\sum_{j=1}^{N}[(g_{1}a_{j}+\Omega _{2})|d_{j}\rangle \langle
a_{j}|e^{i\Delta _{1}^{\prime }t}+(g_{4}a_{j}+\Omega _{3})  \notag \\
&&|e_{j}\rangle \langle c_{j}|)e^{i\Delta _{3}^{\prime
}t}+h.c.]+\sum_{j=1}^{N}J(a_{j}^{\dagger }a_{j+1}+a_{j}a_{j+1}^{\dagger })
\end{eqnarray}

The other two cavity fields denoting with $g_{2}$\bigskip\ and $g_{3}$ in
Eq. (1) will only induce a Stark shift on the level $|b\rangle $ and do not
affect the effective interaction, so we do not write it in Eq. (13). Under
larger detuning $\{|\Delta _{1}^{\prime }|$,\TEXTsymbol{\vert}$\Delta
_{3}^{\prime }|\}\gg $ $\{g_{1}$, $g_{4}$, $\Omega _{2}$ $\Omega _{3}$, $J\}$%
, we adiabatic eliminate atomic excited state and have%
\begin{eqnarray}
H &=&-\sum_{j=1}^{N}\{\frac{g_{1}^{2}}{\Delta _{1}^{\prime }}[|a_{j}\rangle
\langle a_{j}|+|c_{j}\rangle \langle c_{j}|]a_{j}^{\dagger }a_{j}-\nu
_{j}b_{j}^{\dagger }b_{j}\}  \notag \\
&&-\sum_{j=1}^{N}[\frac{\Omega _{3}^{2}}{\Delta _{3}^{\prime }}|c_{j}\rangle
\langle c_{j}|+\frac{\Omega _{2}^{2}}{\Delta _{1}^{\prime }}|a_{j}\rangle
\langle a_{j}|]  \notag \\
&&-\sum_{j=1}^{N}[\frac{\Omega _{2}g_{1}}{\Delta _{1}^{\prime }}%
(a_{j}S_{zj}^{{}}+h.c.),
\end{eqnarray}%
where we need the condition $\frac{g_{1}^{2}}{\Delta _{1}^{\prime }}=\frac{%
g_{4}^{2}}{\Delta _{3}^{\prime }}$, $\frac{\Omega _{3}g_{4}}{\Delta
_{3}^{\prime }}=-\frac{\Omega _{2}g_{1}}{\Delta _{1}^{\prime }}$. Switch
into interaction picture and then adiabatic eliminate once more under the
condition $|\frac{\Omega _{2}g_{1}}{\Delta _{1}^{\prime }}|\ll |\frac{%
g_{1}^{2}}{\Delta _{1}^{\prime }}-\nu _{k}|$. Finally, we have the effective
Hamiltonian 
\begin{equation}
H_{zz}=\sum_{j=1}^{N}\frac{\Omega _{2}^{2}g_{1}^{2}}{\Delta _{1}^{\prime 2}u}%
S_{zj}^{2}+\frac{2J\Omega _{2}^{2}g_{1}^{2}}{\Delta _{1}^{\prime 2}u^{2}}%
S_{zj}S_{zj+1}.
\end{equation}%
where $u=\frac{g_{1}^{2}}{\Delta _{1}^{\prime }}$. Now, the effective
Hamiltonian is $S_{z}S_{z}$ interaction.

During the process of deduction of effective Hamiltonian Eqs.(11) and (15),
we have changed working picture for two times. But the atomic population
probability do not change with the changing of the picture. So, we can
compare the atomic population probability in different picture, for example
we have done in Fig. 2. However, the Hamiltonian Eqs. (13) and (15) do not
affect atomic population probability. Therefore, we do not plot it again.

The two Hamiltonians Eqs. (11) and (15) can be combined into one effective
Hamiltonian if we employ the method proposed by \cite{hartmann2}. The lasers
that generate the Hamiltonian (11) are turned on for a short time interval $%
dt$ ($\Vert H_{xy}\Vert dt\ll 1$) followed by another time interval $dt$ ($%
\Vert H_{zz}\Vert dt\ll 1$) with the lasers that generate the Hamiltonian
(15) turned on. By repeating this sequence until the total time range to be
simulated is covered. The effective Hamiltonian $H=H_{xy}+H_{zz}$ finally
can be obtained.

The decoherence of the system mainly results from the decay mechanisms via
the photons or the excited state $|e\rangle $ ($|d\rangle $). To overcome
the decoherence, the coefficients of effective Hamiltonian $A$, $B$, and $C$
should be much larger than the decays rates of photons and the excited state 
$|e\rangle $ ($|d\rangle $). Using the group of the parameters of Fig. 2
(all the parameters are scaled in $g_{1}$), we have $A=-0.0128$, $B=0.0210$, 
$C=0.0113$. For a strongly coupled single quantum dot--cavity system, $\frac{%
g}{\kappa }\sim 1800$ ($\kappa \sim 1/1800$), $\frac{g}{\gamma }\sim 300$ ($%
\gamma \sim 1/300$) in which $\kappa $ ($\gamma $) means the decay of the
cavity ( the excited state ) have been achieved for off resonance \cite%
{nature}. Therefore $\{A,B,C\}\ll \{\kappa $, $\gamma \}$ can be realized in
experiment.

\section{Conclusion}

We consider a coupled array of $N$ identical cavities, each of which
contains a five-level atom. We show that under large detuning case the atoms
via the exchange of virtual photons can be effectively equal to spin 1
Heisenberg model. Although the coefficients are related one another, we
still can tune them by controlling the laser fields so that our system is a
good simulator for spin 1 Heisenberg model. When operated in a
two-dimensional array of cavities the device is thus able to simulate spin
lattices.

Acknowledgments: The project was supported by NSFC under Grant No.10774020,
and also supported by SRF for ROCS, SEM.

\end{document}